\documentclass[10pt,amsmath,amssymb,nofootinbib,twoside,twocolumn,superscriptaddress,floats,floatfix,aps,prd,preprintnumbers]{revtex4-2}
\usepackage{graphicx}
\usepackage{aas_macros}
\usepackage{amsmath}
\usepackage{epsfig}
\usepackage{soul}
\usepackage[latin1]{inputenc}
\usepackage[english]{babel}
\usepackage[T1]{fontenc}
\usepackage{amssymb}
\usepackage{amsfonts}
\usepackage{hyperref}
\usepackage{epsfig}
\usepackage{colordvi}
\usepackage{psfrag}
\usepackage{color}
\usepackage{dcolumn}
\usepackage{multirow}
\usepackage{epstopdf}
\usepackage{subfig}
\usepackage{caption}
\usepackage{bm}
\usepackage{times}
\usepackage{xcolor}
\usepackage[normalem]{ulem}
\definecolor{ao(english)}{rgb}{0.0, 0.5, 0.0}
\hypersetup{
  colorlinks=true,        
  linkcolor=blue,         
  citecolor=magenta,      
}
\usepackage{enumitem}
\usepackage{dcolumn}
\usepackage{bm}
\usepackage{color}

\begin{document}
\title{Gravitational signatures beyond Newton: exploring hierarchical three-body dynamics}
\author{Pietro Farina}
\email{pietro.farina@unina.it}
\affiliation{Universit\'a di Napoli "Federico II", Compl.~Univ.~di Monte S.~Angelo, Edificio G, Via Cinthia, I-80126, Napoli, Italy}
\affiliation{INFN Sezione di Napoli, Compl.~Univ.~di Monte S.~Angelo, Edificio~G, Via Cinthia, I-80126, Napoli, Italy}

\author{Mariafelicia De Laurentis}
\email{mariafelicia.delaurentis@unina.it}
\affiliation{Universit\'a di Napoli "Federico II", Compl.~Univ.~di Monte S.~Angelo, Edificio G, Via Cinthia, I-80126, Napoli, Italy}
\affiliation{INFN Sezione di Napoli, Compl.~Univ.~di Monte S.~Angelo, Edificio~G, Via Cinthia, I-80126, Napoli, Italy}

\author{Hideki Asada}
\email{asada@hirosaki-u.ac.jp}
\affiliation{Graduate School of Science and Technology, Hirosaki University, Hirosaki 036-8561, Japan}

\author{Ivan De Martino}
\email{ivan.demartino@usal.es}
\affiliation{{Departamento de F\'isica Fundamental, Universidad de Salamanca, Plaza de la Merced, s/n, E-37008 Salamanca, Spain}}
\affiliation{{Instituto Universitario de F\'isica Fundamental y Matem\'aticas, Universidad de Salamanca, Plaza de la Merced, s/n, E-37008 Salamanca, Spain}}

\author{Riccardo Della Monica}
\email{rdellamonica@usal.es}
\affiliation{{Departamento de F\'isica Fundamental, Universidad de Salamanca, Plaza de la Merced, s/n, E-37008 Salamanca, Spain}}

\date{\today}

\begin{abstract}
Hierarchical three-body systems offer a compelling framework to explore the subtle interplay between Newtonian and relativistic gravitational effects in astrophysical environments. In this work, we investigate post-Newtonian corrections to the periastron shift within such systems, focusing on the impact of orbital eccentricity. Modeling the secondary body's influence as a quadrupolar perturbation, we compare Newtonian, Schwarzschild, and post-Newtonian quadrupolar contributions to orbital precession. Our analysis demonstrates that Newtonian quadrupolar effects could be observable, for a long monitoring time, in the orbit of the S87 star around Sagittarius A* if an intermediate-mass black hole is present, under the assumptions of our model. Additionally, post-Newtonian quadrupolar corrections may influence the dynamics of small Solar System bodies in the presence of massive companions.
Although the predicted effects are minute and require long monitoring periods to be measurable, our analysis clarifies how relativistic corrections enter the dynamics of the third body and outlines the conditions under which future observations could reveal them.

\end{abstract}

\maketitle

\section{Introduction}

Self-gravitating three-body systems are very common in many contexts in nature, especially in astrophysics. The classical approach for studying three-body systems deals with the motion of three celestial bodies under mutual gravitational attraction. Throughout the centuries, there have been many attempts to solve this problem until Poincar\'e \cite{poincare} demonstrated that it is intrinsically chaotic. This feature has been confirmed through numerical simulations \cite{burrau, szebehely, szebehely2, agekyan} and also using an analytical statistical mechanics approach, providing a suitable definition of an ergodic three-body system \cite{monaghanI, monaghanII, valtonen, heggie}.

Several attempts were made to study particular configurations in which the dynamics of the system can be predicted analytically. For instance, the case of the equilateral triangular configurations, where the masses are at the vertex of an equilateral triangle at any time (a nonchaotic dynamical evolution of the triangular three-body may also be a source of the emission of  gravitational waves \cite{PhysRevD.93.084027, Yamada15}), or the case of the restricted three-body problem, where the third body has a negligible mass compared to others \cite{valtonen, Yamada15}. Another interesting configuration that can be found in several astrophysical contexts \cite{tokovinin, tokovinin2} and does not show a chaotic evolution is the hierarchical configuration. In this case, the system is composed of a binary system and a more distant third body enabling, in such a way, the application of perturbative methods to investigate the impact of the massive third body on binary systems. This approach, to quadrupole order approximation, led to the discovery of the so-called von Ziepel-Kozai-Lidov (ZKL) perturbations \cite{kozai, ito}, i.e. oscillation in binary's and third body's orbital eccentricities due to the relative inclination of the two orbits. Octupole effects, in this context, give the eccentric Kozai-Lidov (eKL) perturbations involving orbital flips and chaotic dynamics \cite{naoz11}. ZKL and eKL  have been explored in the post-Newtonian (PN) approximation \cite{desitter,droste,fock,damour,willbook}, showing that eKL mechanisms can be inhibited by relativistic periastron precession \cite{naoz2, naoz3}, when
the timescales associated with the 1PN effects are much shorter than the timescales associated with the eKL perturbations. 
Recently, in \cite{archibald}, numerical simulations of a hierarchical three-body system adopting a parametrized PN approximation have been used to probe the equivalence principle. Making a comparison between the theoretical prediction of the post-Keplerian parameters and its observational counterpart from the system PSR J0337+1715, a system made of a millisecond radio pulsars and a white dwarf forming the inner binary system, and an additional outer white dwarf, they improve about three orders of magnitude the bounds of the Nordtvedt parameter \cite{nordtvedt, nordtvedt2}.

Nowadays, interest in studying three-body systems is fueled by the need to deepen our understanding of the environment surrounding supermassive black holes (SMBH). These are black holes with a mass greater than $10^5 M_\odot$ whose existence is observationally confirmed at the center of almost every massive galaxy \cite{ferrarese}. These objects are often surrounded by orbiting gas and clusters of stars offering a natural laboratory for investigating $N$-body systems.  This kind of system may naturally form, therefore,  a hierarchical three-body system whose dynamics can be studied through a PN approximation. For example, galaxy merging may result in the formation of SMBH binaries on a time scale less than the Hubble time due to dynamical friction of the stellar environment \cite{begelman, callegari}. In fact, the time variability of the light curve of the blazar OJ 287 points to the presence of a SMBH binary at its center \cite{sillanpaa, komossa}, providing an observational confirmation of the presence of a SMBH binary at the center of galaxies.

Regarding this paradigm, the presence of Intermediate Mass Black Holes (IMBHs) in the center of galaxies has been explored, for example \cite{lockman}, where the authors found that the presence of IMBHs in the center of Milky Way agrees with flattened cusp in the density profile. Recent observations from \href{https://www.virgo-gw.eu}{\color{blue} VIRGO}, \href{https://www.ligo.caltech.edu/}{\color{blue} LIGO}, and \href{https://gwpo.nao.ac.jp/en/research/kagra.html}{\color{blue} KAGRA} have provided crucial insights on the potential existence of IMBHs studying the gravitational wave emission from the source GW 150914 at redshift $z=0.09$ \cite{ligo} made of two black holes with inferred masses of $\sim 36 M_\odot$ and $\sim 29 M_\odot$.

Emission of the gravitational waves could occur due to both mechanisms the ZKL \cite{amaro-seoane, camilloni} and the dynamical friction generated from a close encounter that could induce the in-spiral phase in a stalled binary system. Some authors have addressed the problem of secular modifications to the orbit of a test particle that forms a binary system with another mass around a SMBH including tidal effects \cite{camilloni2, grilli} finding that those modifications could be relevant in a strong field case.

Here, we will implement the calculations performed in \cite{yamada} of the 1PN contributions to the periastron shift of a test particle due to the presence of a binary system, and we will apply them to different three-body systems to forecast the amplitude of the PN effects. In Sec. \ref{sec:av_metric} we summarize previous work \cite{yamada} and extend it to eccentric binary systems. In Sec. \ref{sec:periad} we focus on obtaining the contributions to periastron shift due to 1PN quadrupolar corrections investigating which terms dominate the dynamics of the three-body system.  Finally, in Sec. \ref{sec:res} we apply our results to toy models describing different astrophysical situations in order to understand how the order of magnitude of the contributions changes, and in Sec. \ref{sec:discussion} we discuss our results and give our final conclusions.

\section{Post-Newtonian averaged metric}
\label{sec:av_metric}

The main aim is to write down an effective averaged metric to treat a hierarchical three-body system in order to compute the equations of motion. These will be used to compute the periastron shift of the third body, which will be assumed to be a test particle. The starting point is the 1PN metric given for an $N$-body system \cite{landau},
\begin{align}
    \label{eqn:pn_metric}
    ds^2 &= \Bigg[-1 + \frac{2G}{c^2} \sum_a \frac{m_a}{|\mathbf{r}-\mathbf{r}_a|}\notag \\
    &\quad -\frac{2 G^2}{c^4} \left(\sum_a \frac{m_a}{|\mathbf{r}-\mathbf{r}_a|}\right)^2 + \frac{3G}{c^4} \sum_a\frac{m_a v_a^2}{|\mathbf{r}-\mathbf{r}_a|} \notag \\
    &\quad -\frac{2G^2}{c^4}\sum_a \sum_{b \ne a}\frac{m_a m_b}{|\mathbf{r}-\mathbf{r}_a||\mathbf{r}_a-\mathbf{r}_b|}\Bigg]c^2 dt^2 \notag \\
    &-\frac{G}{c^3} \sum_a \frac{m_a}{|\mathbf{r}-\mathbf{r}_a|}\left[7 v_{ai}+(\textbf{v}_a \cdot \textbf{n}_a) n_{ai}\right]cdt dx^i  \notag \\
    & \quad+ \left(1+ \frac{2 G}{c^2} \sum_a \frac{m_a}{|\mathbf{r}-\mathbf{r}_a|}\right) \delta_{ij}d x^i dx^j.
\end{align}

Then, following the procedure highlighted in \cite{yamada}, we assume that the three-body system is coplanar and hierarchical which implies that the third body has a semimajor axis $a$ much larger than the one of the internal binary system, $a_B$. Hereby, the subscript $B$ will always refer to the binary parameters.  Additionally, we assume that $m_1 \gg m_2 \gg m_3$, where $m_1$ and $m_2$ are the masses of the two components of the binary system and $m_3$ is the mass of the test particle. Since the three bodies orbit on the same plane we can neglect secular variations of the inclination, eccentricity and longitude of ascending nodes that could affect the periastron shift. Finally, at 1PN quadrupolar order, we can neglect the mixed term in Eq. \eqref{eqn:pn_metric}, which gives rise to an additional contribution to the periastron shift that matches the contribution from the binary orbital angular momentum \cite{landau}, $L_B$, which is given by
\begin{equation}
    \label{eqn:Lcontr}
    \dot{\varpi}_L=-\frac{4Gm_2\sqrt{a_B(1-e_B^2)}}{c^2a^{3/2}(1-e^2)^{3/2}}n,
\end{equation}
where $a_B$ and $e_B$ are the semimajor axis and the eccentricity of the binary, respectively, $a$ and $e$ are the semimajor axis and the eccentricity of the third body, respectively, and we have introduced the mean motion $n=2\pi/P$, being $P$ the orbital period of the third body.

The next step consists in doing a multipole expansion of the quantity $1/|\mathbf{r}-\mathbf{r}_a|$ in the metric \eqref{eqn:pn_metric} and rewriting it in the center of mass reference frame. We expand up to quadrupole order, working in the context of ZKL effects. To do so, since we are considering that the binary system has an eccentricity different from zero, we have to consider the 1PN correction to the center of mass given in \cite{damour},
\begin{equation}
    \textbf{r}_{1,2}=\left[\pm \frac{\mu}{m_{1,2}}+ \frac{\mu (m_1-m_2)}{2 M c^2}\left(V^2- \frac{GM}{R}\right)\right]\textbf{R},
\end{equation}
where $\mathbf{R}=\mathbf{r}_1-\mathbf{r}_2$, $\mathbf{V}=\mathbf{v}_1-\mathbf{v}_2$, $M$ is the total binary mass and $\mu$ is its reduced mass.
Taking all those steps, the metric can be recast as,
\begin{align}
    ds^2&=\biggl\{-1+ \frac{2GM}{c^2r}-\frac{G^2 m_2 M}{c^4 r^2}\left(5-\frac{4 R}{a_B}\right) \mathbf{N}\cdot\mathbf{n} \notag \\
    &-\frac{Gm_2R^2}{c^2r^3}\left[1-3(\mathbf{N}\cdot\mathbf{n})^2\right]-\frac{2G^2M^2}{c^4r^2} \notag \\
    &+\frac{2G^2m_2MR^2}{c^4r^4}\left[1-3(\mathbf{N}\cdot\mathbf{n})^2\right]\notag \\
    &+\frac{G^2Mm_2}{c^4 rR}\left(2-\frac{3R}{a_B}\right)\biggr\}c^2dt^2\notag \\
    &+\biggl\{1+\frac{2GM}{c^2r}-\frac{Gm_2R^2}{c^2r^3}\left[1-3(\mathbf{N}\cdot\mathbf{n})^2\right]\biggr\}\delta_{ij}dx^i dx^j,
\end{align}

where $\mathbf{N}=\mathbf{R}/R$, $\mathbf{n}=\mathbf{r}/r$ and we used $m_1\simeq M$ and $m_2\simeq \mu$. Henceforth, we shall use the polar coordinates $(r, \theta, \phi)$ as a reference frame.
The following step is to introduce the averaged quantities in the metric, where the average is taken over the binary period ($P_B$). Generally speaking, taking a physical magnitude $x$ which depends on time and varies over the orbit, its averaged value can be written as:
\begin{align}
    \langle x \rangle_t &=\frac{1}{P_B} \int_0^{P_B} x(t) dt=\notag \\
    &=\frac{1}{2 \pi a_B^2 \sqrt{1-e_B}} \int_0^{2 \pi} x(\varphi) R^2(\varphi) d \varphi.
\end{align}
where $\varphi$ is the true anomaly of the binary system. The above equation differs from the corresponding one in \cite{yamada} because of the introduction of the eccentricity which gives rise to the term $R^2(\varphi)/a_B^2 \sqrt{1-e_B}$. Finally, to obtain a Schwarzschild-like 1PN metric, we need also to carry out the following change of coordinates:
\begin{equation}
    \label{eqn:iso_transf}
    \rho^2=\left(1 + \frac{2GM}{c^2 r}+ \frac{Gm_2a_B^2}{c^2r^3} \mathcal{B}\right) r^2,
\end{equation}
where $\rho$ is the isotropic radial coordinate, and we have defined the quantity,
\begin{align}
    \mathcal{B}&=\frac{3}{2}\big[(1+4e_B^2)\cos^2(\phi-\omega_B)\notag \\
    &\quad +(1-e_B^2)\sin^2(\phi-\omega_B)\big]-1-\frac{3}{2}e_B^2,
\end{align}
where $\omega_B$ is the argument of periastron of the binary system.
Thus, we obtain,
\begin{align}
    \label{eqn:av_metric}
    \langle ds^2 \rangle&=\Bigg[-1 +\frac{r_S}{\rho}\left(1- \frac{Q}{2a_B^3}\right)- \frac{r_S Q}{\rho^2 a_B^2} \mathcal{A}+ \frac{Q}{\rho^3}\mathcal{B} \notag \\
    &+\frac{r_S Q}{\rho^4} \mathcal{B}\Bigg]c^2 dt^2+\left(1+ \frac{r_S}{\rho}+ \frac{Q}{\rho^3}\mathcal{B}\right)d \rho^2 \notag \\
    &+\rho^2d\phi^2,
\end{align}
where $r_S=2GM/c^2$ is the effective Schwarzschild radius of the binary, and,
\begin{equation}
    Q= \frac{Gm_2a_B^2}{c^2}, \qquad \mathcal{A}=e_B \cos(\phi-\omega_B),
\end{equation} 
is the moment induced by body with mass $m_2$. It is worth note that taking $e_B=0$ recovers the metric in Eq. (4) of \cite{yamada}. 

\section{Equation of motion and periastron shift}
\label{sec:periad}

Starting from the effective metric in Eq. \eqref{eqn:av_metric}, we find the effective Lagrangian for the test particle,
\begin{align}  
    \label{eqn:eff_lagr}
    \mathcal{L}&=\Bigg[-1 +\frac{r_S}{\rho}\left(1- \frac{Q}{a_B^2}\right)- \frac{r_S Q}{\rho^2 a_B^2} \mathcal{A}+ \frac{Q}{\rho^3}\mathcal{B} \notag \\
    &+\frac{r_S Q}{\rho^4} \mathcal{B}\Bigg]c^2 \dot{t}^2+\left(1+ \frac{r_S}{\rho}+ \frac{Q}{\rho^3}\mathcal{B}\right)\dot{\rho}^2\notag \\
    &+\rho^2\dot{\phi}^2,
\end{align}
Obviously, $t$ and $\phi$ are cyclic; consequently, we can write the constants of the motion as:
\begin{align}   
   \frac{\partial\mathcal{L}}{\partial \dot{t}} \equiv \mathcal{E}&=\biggl[1 -\frac{r_S}{\rho}\left(1- \frac{Q}{a_B^2}\right)+\frac{r_S Q}{ \rho^2 a_B^2} \mathcal{A}- \frac{Q}{\rho^3}\mathcal{B}\nonumber\\
     & - \frac{r_S Q}{\rho^4} \mathcal{B}\biggr] \dot{t}, \label{eqn:const_mot1} \\
     \label{eqn:const_mot2} \frac{\partial\mathcal{L}}{\partial \dot{\phi}} \equiv \ell&= \rho^2 \dot{\phi}.
\end{align}
Here, $\mathcal{E}$ is the dimensionless energy,
\begin{equation}
    \mathcal{E}=E/mc^2,
\end{equation} 
being $E$ the energy of the third body, and $\ell$ the specific angular momentum,
\begin{equation}
    \ell=L/mc,
\end{equation}
where $L$ is the third body's angular momentum.

Using Eqs. \eqref{eqn:const_mot1}, \eqref{eqn:const_mot2}, and \eqref{eqn:eff_lagr}, we find the following expression:
\begin{align}
    \label{eqn:energy_int}
    &\left(1+\frac{2Q}{\rho^3} \mathcal{B}\right) \frac{\dot{\rho}^2}{c^2} =\,\mathcal{E}^2-1+\frac{r_S}{\rho}\left(1- \frac{Q}{2a_B^3}\right)- \frac{r_S Q}{2 \rho^2 a_B^2}\mathcal{A} \notag \\ 
    &\qquad+\frac{Q}{\rho^3}\mathcal{B}+\frac{r_SQ}{\rho^4}\mathcal{B}-\left(1-\frac{r_S}{\rho}-\frac{Q}{\rho^3}\right)\frac{\ell^2}{c^2 \rho^2}.
\end{align}
To compute the periastron shift $\Delta \omega$ we have to solve the following integral \cite{chandra, weinberg}:
\begin{equation}
    \label{eqn:deltaom}
    \Delta \omega=\int_{\rho_a}^{\rho_p}d\phi+\int_{\rho_p}^{\rho_a}d\phi- 2 \pi,
\end{equation}
where $\rho_a$ and $\rho_p$ are the isotropic coordinates of the apoastron and periastron, respectively. Before carrying out the integration, it is convenient to make a change of  variable $u=1/\rho$ to write the Eq. \eqref{eqn:energy_int} in the form,
\begin{equation}
    \left(\frac{du}{d\phi}\right)^2= f(u).
\end{equation}

Inserting the previous equation in \eqref{eqn:deltaom}, and dividing both members of the latter by period of the third body, $P$, we find the following contributions to the periastron shift rate:
\begin{align}
    \dot{\varpi}_Q&=\frac{3Gm_2a_B^2}{8c^2a^3(1-e^2)^3}\bigg\{16+9e^2+ e_B^2\Big[24+\notag \\
     &\quad+\frac{e^2}{4}(65\cos 2 \delta\omega+54)\Big]\bigg\}n \label{eqn:contr1}\\
    \label{eqn:contr2} \dot{\varpi}_N&= \frac{3m_2a_B^2}{8Ma^2(1-e^2)^2}(2+3e_B^2)n,
\end{align}
Equation \eqref{eqn:contr1} is the 1PN contribution at quadrupole order,  while the Eq. \eqref{eqn:contr2} shows the Newtonian contribution at quadrupole order. The latter is mathematically equivalent to the quadrupolar Newtonian contribution of an oblate source \cite{misner}. Additionally, we stress that the Newtonian contribution is the same as that found in \cite{willhex} in the test particle approximation, for a binary system with $m_2 \ll m_1$. 

\begin{figure*}
    \centering
    \includegraphics[scale=0.52]{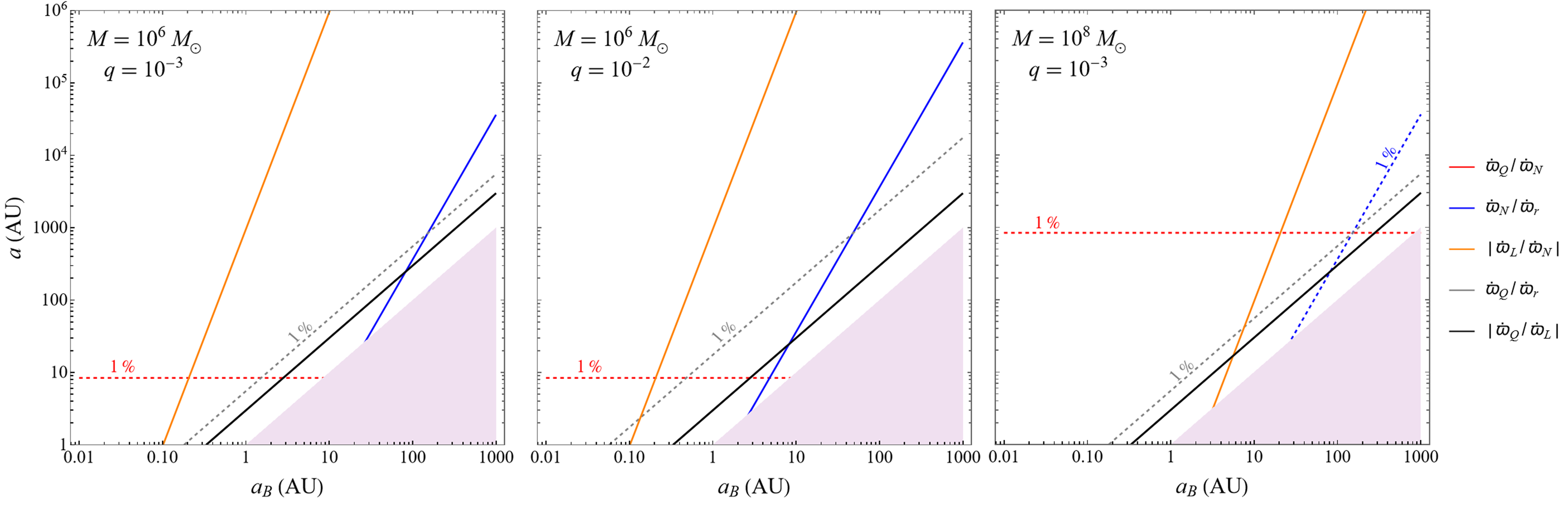}
    \caption{Ratios between the different contributions: red line for $\dot{\varpi}_Q/\dot{\varpi}_N$, blue line  line for $\dot{\varpi}_N/\dot{\varpi}_r$, orange line for  $|\dot{\varpi}_L/\dot{\varpi}_N|$, gray line for $\dot{\varpi}_Q/\dot{\varpi}_r$ and black line for  $|\dot{\varpi}_Q/\dot{\varpi}_L|$. Solid lines indicates the points ($a_B$, $a$) for which the ratio is equal to 1, dashed lines correspond to a ratio equal to 0.01. In the region above the solid lines the numerator become subdominant. In the region above the dashed lines the numerator contribution becomes less than 1\%. The shaded purple region is excluded, giving $a<a_B$. In the left and central panels the total mass of the system is $M=10^6\, M_{\odot}$, while in the right panel $M=10^8\, M_{\odot}$. The mass ratio $q=m_2/M$ is $10^{-3}$ (left and right panel) and $10^{-2}$ (central panel). For each panel $e=0.2$ and $e_B=0.5$.}
    \label{fig:comptog}
\end{figure*}

Now we can compare the orders of magnitude of each contribution. Figure \ref{fig:comptog} shows the ratios between them for various total masses and mass ratios $q=m_2/M$. Each color corresponds to a specific ratio: red for $\dot{\varpi}_Q/\dot{\varpi}_N$, blue for $\dot{\varpi}_N/\dot{\varpi}_r$, orange for  $|\dot{\varpi}_L/\dot{\varpi}_N|$, gray for  $\dot{\varpi}_Q/\dot{\varpi}_r$ and black for  $|\dot{\varpi}_Q/\dot{\varpi}_L|$. Solid lines indicate the combinations of $a$ and $a_B$ where a given ratio is equal to one, while dashed lines correspond to a ratio of 0.01. In regions above the solid lines the numerator is the subdominant component, while for region above the dashed lines the numerator contributes less than $1\%$. In all panels, the orbital eccentricity of the third body is fixed to 0.2 while $e_B=0.5$. Varying the latter parameter does not significantly change the results shown in Fig. \ref{fig:comptog}. The shaded purple region is forbidden since $a<a_B$ in it. The ratio $\dot{\varpi}_L/\dot{\varpi}_r$ is not shown, as it remains of the order $10^{-6}$. In each case, $\dot{\varpi}_Q$ represents a maximum of $\sim (5\div 6)\%$ with respect to both $\dot{\varpi}_N$ and $\dot{\varpi}_r$ when $a_B \sim a$. A similar behavior is observed for $\dot{\varpi}_N/\dot{\varpi}_r$, when the central mass is a SMBH with $M=10^8\; M_{\odot}$, where the relativistic term dominates the periastron advance. The contributions sensitive to the increase $q$ are those involving $\dot{\varpi}_r$ (blue lines and gray lines in Fig. \ref{fig:comptog}), which does not depend on $m_2$.

\subsection{Toy models}
\label{ssec:toy}

\begin{table}
\centering
\setlength{\tabcolsep}{3.5pt}
\renewcommand{\arraystretch}{1.5}
\begin{tabular}{c|cccccc}
\hline
        \# Toy  & $m_1\;(M_{\odot})$ & $m_2 \;(M_{\odot})$ & $a_B$ (AU) & $e_B$ & $a$ (AU) & $e$\\\hline
        $1$ & $30$ & $[10^{-4},10^{-2}]$ & $[10,100]$ & $0.6$ & $3\cdot10^3$ & $0.2$\\
        $2$ & $10^6$ & $10$ & $[10,100]$ & $[0,1)$ & $3\cdot10^3$ & $0.2$\\
        $3$ & $10^9$ & $[10^3,10^6]$ & $10^4$ & $[0,1)$ & $10^6$ & $0.2$\\
        \hline
\end{tabular}
\caption{Parameter used for each toy model, where different couples of parameters are chosen as variables. In that case, we indicated the ranges we will consider.}
\label{tab:toys} 
\end{table} 

To explore the parameter space and investigate in which cases the different contributions to the periastron shift became dominant, we first explore three possible systems that might serve as a test bench, whose main orbital parameters are listed in Table \ref{tab:toys}. 
In this preliminary analysis we will consider only the order of magnitude of $\dot{\varpi}_Q\sim 3Gm_2a_B^2(16+9e^2)(2+3e_B^2)/[8c^2a^3(1-e^2)^3]$.

\begin{figure*}
    \centering
    \includegraphics[scale=0.52]{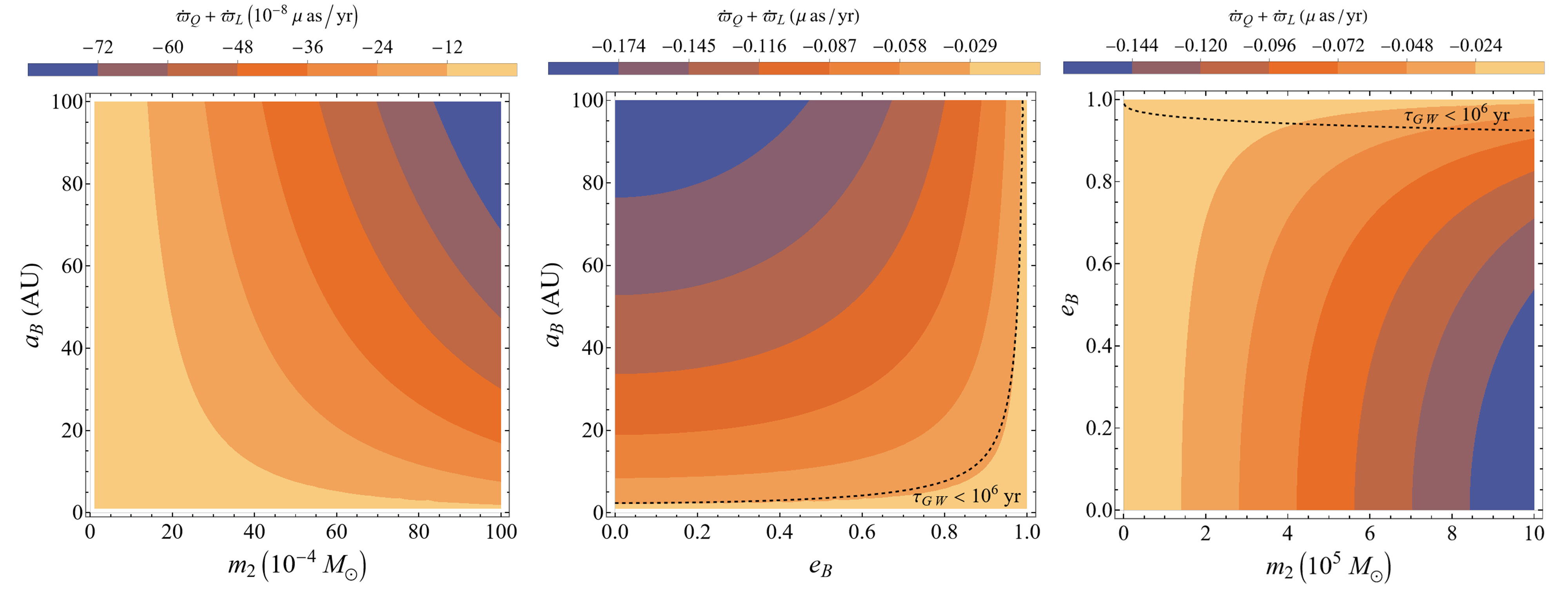}
    \caption{Periastron shift rate related to the quadrupolar and the orbital momentum terms. The Newtonian component is not considered since it would be the dominant one. The left, middle, and right panels are particularized for the toy models 1, 2, and 3 respectively. In the panels we depict how the periastron shift rate $\dot{\varpi}_Q+\dot{\varpi}_L$ depends on the specific pair of parameters that are varied according to Table \ref{tab:toys}. The dashed black line in the central and right panels represents the parameters that give a gravitational-wave damping timescale equal to 1 Myr.}
    \label{fig:toy}
\end{figure*}

The absolute value of the periastron shift rate increases with higher values of the semimajor axis $a_B$ and the mass $m_2$ of the binary system. This is quite obvious because each rate $\dot{\varpi}_i$ is proportional to some power of these two variables. Therefore, increasing $a_B$ makes the quadrupolar effect dominant because this means that the celestial body corresponding to the mass $m_2$ is closer to the third body. In Fig. \ref{fig:toy} we show the magnitude of the periastron shift rate related to the quadrupolar component and the orbital angular momentum. We always avoid plotting the Newtonian component, since it would be the dominant one. The left, middle, and right panels are particularized for the toy models 1, 2, and 3 respectively. In the panels we depict how the periastron shift rate $\dot{\varpi}_Q+\dot{\varpi}_L$ depends on the specific pair of parameters that are varied according to Table \ref{tab:toys}. Since we are considering relativistic objects, it is convenient to exclude the orbital elements that give a gravitational-wave damping timescale shorter than the binary period; otherwise, we should take the semimajor axis variation into account while integrating over the binary period. A suitable choice for the toy models 2 and 3 is to choose parameters that give gravitational-wave damping timescale $\tau_{GW}>1$ Myr. Such a time scale is depicted in the middle and left panels as a dashed black line.

We notice that for each toy model the periastron shift is always negative, since $\dot{\varpi}_L$ prevails on the quadrupolar contribution, so we are exploring the parameters above the black lines in Fig. \ref{fig:comptog}.
As mentioned in the beginning of this section, the Newtonian contribution is the highest one, bringing a precession rate magnitude of $\sim 2\;\mu$as/yr, for the first toy model. In fact, even  considering the Schwarzschild precession rate,
\begin{equation}
    \label{eqn:prec_sch}
    \dot{\varpi}_r=\frac{3GM}{c^2a(1-e^2)}n,
\end{equation}
since the central mass is not a relativistic object, all the other component are subdominant.

\begin{figure*}
        \includegraphics[scale=0.6]{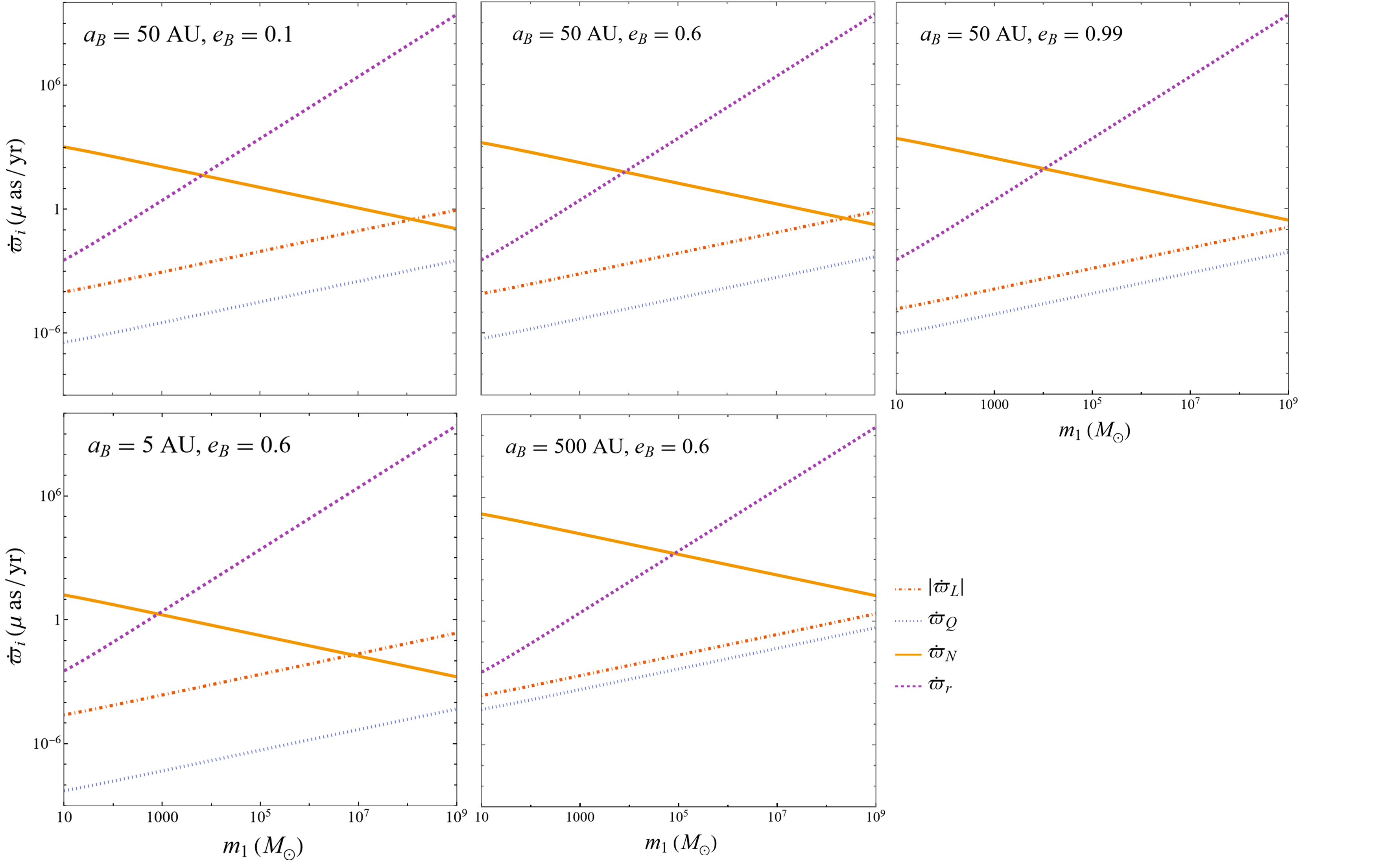}
    \caption{Amplitude of all the different components that contribute to the total precession rate, as a function of the mass $m_1$. We depict the Newtonian contribution in Eq. \eqref{eqn:contr2} as a solid orange line, the relativistic Schwarzschild term in Eq. \eqref{eqn:prec_sch} as a purple dashed line, the quadrupolar contribution in Eq. \eqref{eqn:contr1} as blue filled circles and, finally, the angular momentum term in equation \eqref{eqn:Lcontr} as a red dot-dashed line. The figure is particularized for a system with mass $m_2=2M_\odot$ and the semimajor axis and the eccentricity of the third body $a=3000$ AU and $e=0.2$, respectively. Then, in the upper panels, we fix $a_B = 50$ AU, and vary the eccentricity in the range $e_B=[0.1, 0.99]$. Conversely, in the lower panel, we set $e_B=0.6$ and vary the semimajor axis in the range $a_B=[5, 500]$AU.}
    \label{fig:toy2}
\end{figure*}

The contours for the contributions of the precession rate $\dot{\varpi}_Q+\dot{\varpi}_L$, in the middle panel of Fig. \ref{fig:toy}, show that for large values of both $e_B$ and $a_B$ the value of the periastron shift increases since $\dot{\varpi}_Q$ starts to become relevant but always is smaller than $\dot{\varpi}_L$. The sum of the two contribution becomes larger for values of the eccentricity spanning the range 0 to 0.5 and semimajor axis starting from 60 AU. In the right panel of  Fig. \ref{fig:toy}, we depict the contours for the precession rate $\dot{\varpi}_Q+\dot{\varpi}_L$ as function of the pair of parameters ($e_B, m_2$) for the toy model 3, which is made of two SMBHs. The contribution of $\dot{\varpi}_L$ and $\dot{\varpi}_Q$ remains of the same order of magnitude as the previous toy model (middle panel). This dependence of $\dot{\varpi}_Q$ and $\dot{\varpi}_L $  is due to the fact that these contributions depend linearly on $m_2$, and depend on an inverse power of the semimajor axis, therefore, increasing $m_2$ leads to increasing $a$ and the two effects are balanced.

For clarity, we show in Fig. \ref{fig:toy2} all components that contribute to the total precession rate. We show the Newtonian contribution in Eq. \eqref{eqn:contr2} as a solid yellow line, the relativistic Schwarzschild term in Eq. \eqref{eqn:prec_sch} as a purple dashed line, the quadrupolar contribution in Eq. \eqref{eqn:contr2} as blue filled circles and, finally, the angular momentum term in Eq. \eqref{eqn:Lcontr} as a red dot-dashed line. The gravitational wave damping time scale is not plotted since, for the specific choice of the orbital parameters, it resides in a region outside the axis margins. The figure is particularized for a system with mass $m_2=2M_\odot$ and the semimajor axis and the eccentricity of the third body $a=3000$ AU and $e=0.2$, respectively. Then, in the upper panels, we fix $a_B = 50$ AU, and vary the eccentricity in the range $e_B=[0.1, 0.99]$. Conversely, in the lower panel, we set $e_B=0.6$ and vary the semimajor axis in the range $a_B=[5, 500]$AU. Obviously, depending on the systems, the dominant terms are the Newtonian one or the 1PN Schwarzschild precession. Increasing the central mass of the binary system makes the system relativistic, which turns in an angular momentum term that, in some cases, can be higher than the Newtonian contribution, though several orders of magnitude below the 1PN Schwarzschild contribution. For instance, in the case where the semimajor axis and the eccentricity of the binary system are set to $a_B=5$ AU and  $e_B=0.6$, respectively, which is depicted in the lower left panel of the Fig. \ref{fig:toy2}, if the central mass is set to $m_1=10^9M_\odot$, the Newtonian contribution to the precession is $\dot{\varpi}_N \sim 10^{-3} \, \mu$as/yr and the angular momentum one is $\dot{\varpi}_L \sim 0.1 \, \mu$as/yr. While the latter is the largest, it is still well below $\dot{\varpi}_r \sim 10^3 \;\text{as/yr}$.

\section{Observational Test bed}
\label{sec:res}

In the following sections we will apply our model to potential observational cases to determine whether those additional contribution to the precession rate may be observed or not.

\subsection{Hypothetical Intermediate Mass Black Hole in the Galactic Centre}
\label{ssec:IMBH_Sgr}

As mentioned in the Introduction, the hierarchical nature of the galaxy formation paradigm suggests that galaxy mergers may result in the formation of binaries systems made of SMBHs \cite{callegari}. Here,  we want to study whether a putative IMBH in the Galactic Center may be observationally confirmed through the observation of the orbital motion of the S87 star.  The existence of such an IMBH could be due to a minor merger of the Galaxy with a dwarf galaxy or even with a globular cluster \cite{rashkov}. In fact, since IMBHs are a fundamental ingredient in the formation of SMBHs in the centers of galaxies, their presence in the central parsecs of the Milky Way was postulated in \cite{gualandris, will23, naoz1, generozov}.

The gravitational potential of the Galactic Center is dominated by a compact, bright radio source named Sagittarius A* (Sgr A*). Recent technological advances, such as the advent of adaptive optics, have made it possible to observe stars orbiting around SgrA* and to monitor their positions throughout the last three decades (e.g. \cite{ghez, gillessen}). Observations by the Event Horizon Telescope collaboration have provided evidence for the shadow of the black hole \cite{EHTSgrAp1,eht2}. Thus, the proximity of the Milky Way's Galactic Center provides a unique laboratory for addressing issues in the fundamental physics of SMBH \cite{2022JCAP...03..007D,2023PhRvD.107d4038C,2023MNRAS.519.1981D,2023JCAP...08..039F,2023PhRvD.108j1303D,2023PhRvD.108l4054D,2024PhRvD.109b4016D,2024arXiv241022864D,EHTSgrAp6,SVLBGrav2021,2018NatAsmizuno}, their impact on the central regions of galaxies \cite{Mapelli2016}, and their role in galaxy formation and evolution \cite{Bryant2021, Alexander2017}. For a comprehensive review, see \cite{2023RPPh...86j4901D}.

In order to obtain the results shown in Secs. \ref{sec:av_metric} and \ref{sec:periad}, we used a star whose orbital eccentricity is sufficiently small, therefore we chose the S87 star whose orbit has an eccentricity $e=0.224$, a semimajor axis of 22450 AU, and an orbital period of 1640 years \cite{gillessen}. The Keplerian parameters relative to its orbit are summarized in Table \ref{tab:Sparameters}. In the table we also include the epoch of the last periastron passage, $t_P$. 
\begin{table}
\centering
\setlength{\tabcolsep}{4pt}
\renewcommand{\arraystretch}{1.5}
    \begin{tabular}{cccccccc}
    \hline
    Star & $a$ (AU)& $e$ & $i\;(^{\circ})$ & $\Omega\;(^{\circ})$ & $\omega\;(^{\circ})$ & $P$ (yr) & $t_P$ (yr) \\
    \hline
    S87 & 22450 & 0.224 & 119.54 & 106.32 & 336.1 & 1640 & 611 \\
\hline
\end{tabular}
\caption{Keplerian orbital parameters for S87 \cite{gillessen}.}
\label{tab:Sparameters}
\end{table} 
We must also suppose that the triple system made of SgrA*, the putative IMBH and S87 is coplanar. Obviously, the obtained results would be modified if the mutual inclination between the SgrA*-IMBH system and S87 were assumed different from zero. In that case, one has to take into account the contribution due to nodal precession as well as possible secular variations in the inclination and eccentricity that can change the precession rate. These cannot be simply considered to be constants into the quadrupolar relativistic effects that we are taking into consideration in this work, due to the long orbital timescale, which is comparable with the one of the secular effects. 

\begin{figure}
    \centering
    \includegraphics[scale=0.53]{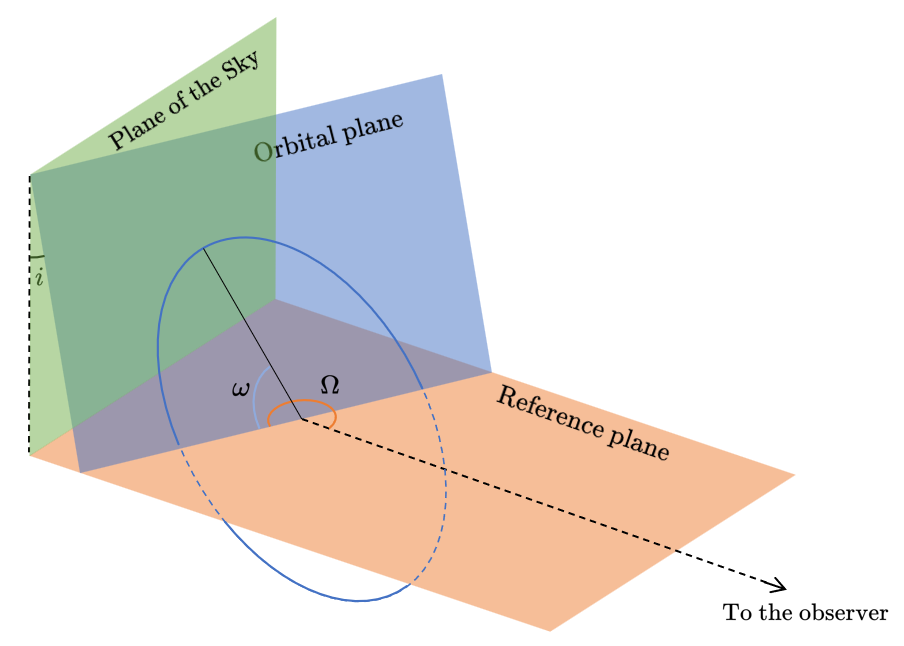}
    \caption{Illustration of the relationship between the plane of the sky, the orbital plane, and the plane of the observer.}
    \label{fig:schema_angoli}
\end{figure}

To forecast whether the quadrupolar effects are observable, we have to use an astrometric model to convert the variation of the argument of the periastron advance on the orbital plane in measurable quantities in the plane tangent to the celestial sphere, i.e. right ascension and declination. In particular we use the astrometric model discussed in \cite{hyman}, in which the right ascension is given by
\begin{equation}
    \label{eqn:transf_ardec}
    \alpha= \frac{1}{d}(Bx+Gy),
\end{equation}
and the declination is 
\begin{equation}
    \label{eqn:transf_ardec2}
    \delta=\frac{1}{d}(Ax+Fy),
\end{equation}
where $d$ is the distance between the observer and the center of mass of the system, $x$ and $y$ are the Cartesian coordinates of the S-Star on the orbital plane, and $A$, $B$, $F$, and $G$ are the Thiele-Innes parameters \cite{oneil,hyman},
\begin{align}
    A&= \cos \Omega \cos \omega-\sin\Omega \sin \omega \cos i, \notag \\
    B&=\sin \Omega \cos \omega + \cos \Omega \sin \omega \cos i, \notag \\
    F&=-\cos \Omega \sin \omega -\sin\Omega \cos \omega \cos i, \notag \\
    G&= -\sin \Omega \sin \omega +\cos \Omega \cos \omega\cos i.
\end{align}
Here, we introduced the inclination angle, $i$, between the orbital plane of S87 and plane of the sky, the argument of periastron $\omega$, and the longitude of the ascending node, $\Omega$. For a graphic representation of these angles see Fig. \ref{fig:schema_angoli}.

Using the previous results, we compute the difference in the right ascension and declination between a Keplerian orbit and one taking into account the Newtonioan and 1PN effects on the advance of the periastron, $| \alpha_K- \alpha_i|$ and $| \delta_K- \delta_i|$, respectively, where the subscript $i$ indicates a generic contribution among those in Eqs. \eqref{eqn:Lcontr}, \eqref{eqn:contr1}, \eqref{eqn:contr2} and their sum (subscript $C$). To include relativistic corrections in the star's position we consider a Keplerian orbit with varying Thiele-Innes parameters, since $\omega$ can change due to the specific correction term taken into account. Regarding the position of the star at the current time, we use the orbital data presented in \cite{gillessen}. Then, we predict future observations using Kepler's equation
\begin{equation}
    M=\xi-e\sin \xi,
\end{equation}
where $M=n \Delta t$ is the mean anomaly and $\xi$ is the eccentric anomaly of the S87's orbit. In particular, we use an iterative method (we refer to Chapter 3 of \cite{valtonen} for further details) to solve the previous equation which, since we are considering small eccentricities, converges very rapidly to an approximate solution.

Firstly, we take into account an observational period starting today and covering 30 years, to evaluate the order of magnitude of the various contributions. Let us start to evaluate the magnitude of the different terms ($\Delta \omega_i$) that would add a contribution to the periastron advance. To evaluate these terms we set the mass of SgrA* to $m_1=4.3 \times 10^6\, M_{\odot}$, the mass of the IMBH to $5\times 10^3 \; M_{\odot}$, the semimajor axis and the eccentricity of the system SgrA*-IMBH to  $a_B=5000$ AU and $e_B=0.5$, respectively. The term $\dot{\varpi}_Q$ gives rise to an additional periastron advance of $\Delta \omega_Q \sim 26\,\mu$as on a timescale of 30 year. In terms of shifts in the right ascension and declination, $\dot{\varpi}_Q$ gives rise to a difference of the order of $10^{-3\div4}\mu$as, respectively, which is well below the  detection threshold of \href{https://www.eso.org/sci/facilities/paranal/instruments/gravity.html}{\color{blue} GRAVITY}, which has an astrometric uncertainty of $\sim 50\, \mu$as \cite{gravity2}. The same can be done with the term $\dot{\varpi}_L$, that gives rise to an additional periastron advance of $\Delta \omega_L \sim 0.9$ mas.  In this case, the difference in the right ascension and declination are $|\alpha_K-\alpha_{L}|\sim 10^{-2}\, \mu$as and $|\delta_K-\delta_{L}|\sim 10^{-3}\,\mu$as, respectively. Again, they are not large enough for a future detection. The Newtonian contribution $\dot{\varpi}_N$ gives rise to an additional periastron advance of $\Delta \omega_N \sim 0.3$ as, which, in terms of the difference in the right ascension and declination, translates to $|\alpha_K-\alpha_N| \sim 0.1$ mas and $|\delta_K-\delta_N|\sim 2\,\mu$as, respectively. The first one is above the detection threshold of GRAVITY. It has to be noted that these differences depend on the star's position along its orbit relative to the observer.

\begin{figure*}
    \includegraphics[scale=0.528]{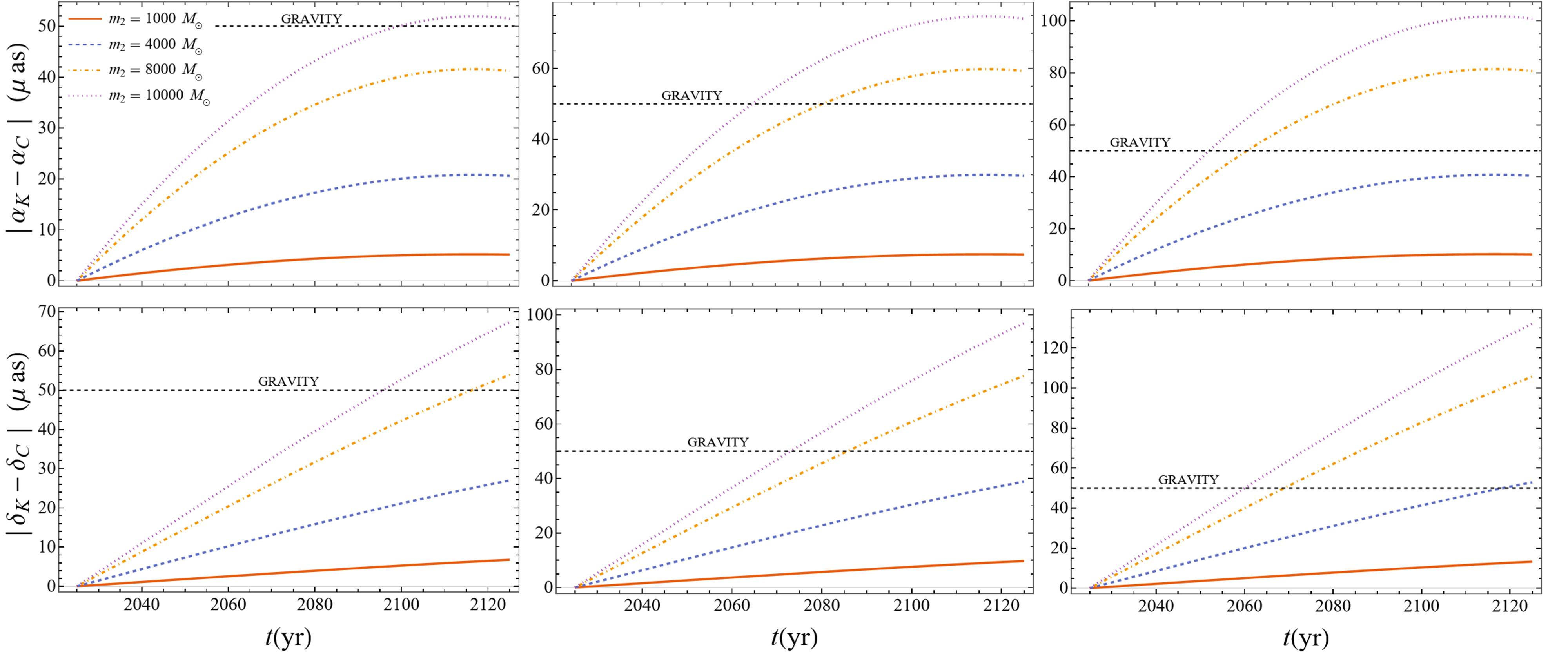}
    \caption{Panels representing the difference in terms of right ascension (\textit{upper panels}) and declination (\textit{lower panels}) between a Keplerian orbit and one considering $ \dot{\varpi}_C=\sum \dot{\varpi}_i$, as a function of time, for S87. The parameters are $M=4.3\cdot10^6 M_{\odot}$, $e_B=0.6$ and, from the left to the right $a_B=5000, \, 6000, \, 7000$ AU. The dashed horizontal black line represents the uncertainty of GRAVITY \cite{gravity2}.}
    \label{fig:S87_time}
\end{figure*}

\begin{figure}
    \centering
    \includegraphics[width=1\linewidth]{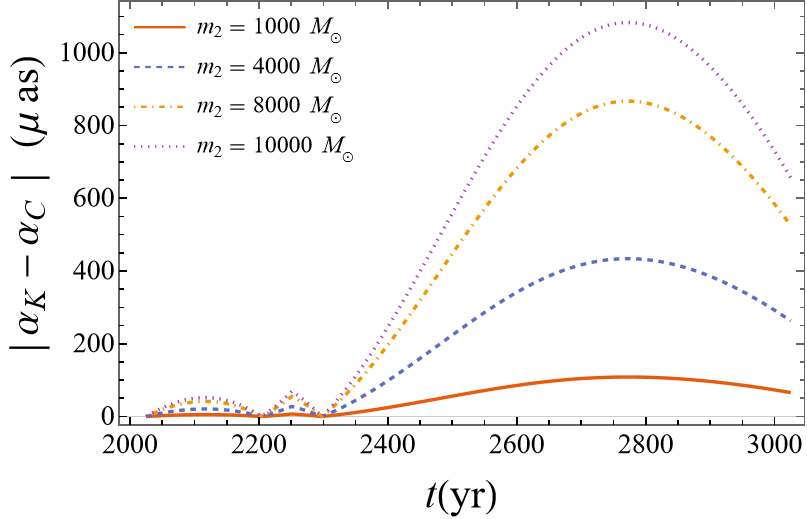}
    \caption{Difference in the right ascension between a Keplerian orbit and the one considering $ \dot{\varpi}_C=\sum\dot{\varpi}_i$. The figure is particularized for a semimajor axis of the SgrA*-IMBH system of $a_B=5000$ AU, and an eccentricity $e_B=0.6$.}
    \label{fig:es1000}
\end{figure}

Given these results, in Fig. \ref{fig:S87_time}, we show the differences between the Keplerian orbit of S87 and the same orbit considering all the additional contributions to the periastron shift, that we have presented so far, in terms of the right ascension and declination, and how they grow with time. The dominant contribution will be the Newtonian one. We set the mass of the IMBH to four different values: $10^3M_\odot$ (red solid line), $4\times10^3M_\odot$ (blue dashed line),  $8\times10^3M_\odot$ (yellow dot dashed line),  and $10^4M_\odot$ (purple dotted line). The dotted black line shows the detection threshold of the current GRAVITY \cite{gravity2} instruments. From the left to the right panel, we set the semimajor axis of the SgrA*-IMBH system to $a_B=5000, \, 6000, \, 7000$ AU, respectively, while the eccentricity is always set to  $e_B=0.6$. The functional form of the resulting profiles does not change, considering different semimajor axes for the SgrA*-IMBH system, because the dominant quantity $\dot{\varpi}_N$ depends linearly on $a_B$ and only changes the magnitude of the contribution at each time.  For instance, considering $a_B=6000$ AU, the difference in right ascension due to an IMBH of $8000\;M_{\odot}$ could be observed by GRAVITY in $\sim 55$ years while the difference in declination would take $\sim 60$ years to be detected. Generally speaking, GRAVITY could eventually detect differences in the right ascension and the declination, only for IMBH masses above $4000M_\odot$. For such IMBHs, only setting the semimajor axis of the SgrA*-IMBH system to $a_B=7000$ AU, the difference in  declination crosses the GRAVITY threshold in $\sim 95$ yr. Finally, in Fig. \ref{fig:es1000} we show how the differences in the right ascension and the declination change on a time span of 1000 yr (comparable with the orbital period), for a specific case with the semimajor axis of the SgrA*-IMBH system set to $a_B=5000$ AU. 

\subsection{Solar System}
\label{ssec:solar_system}
\begin{figure*}
    \includegraphics[width=1.7\columnwidth]{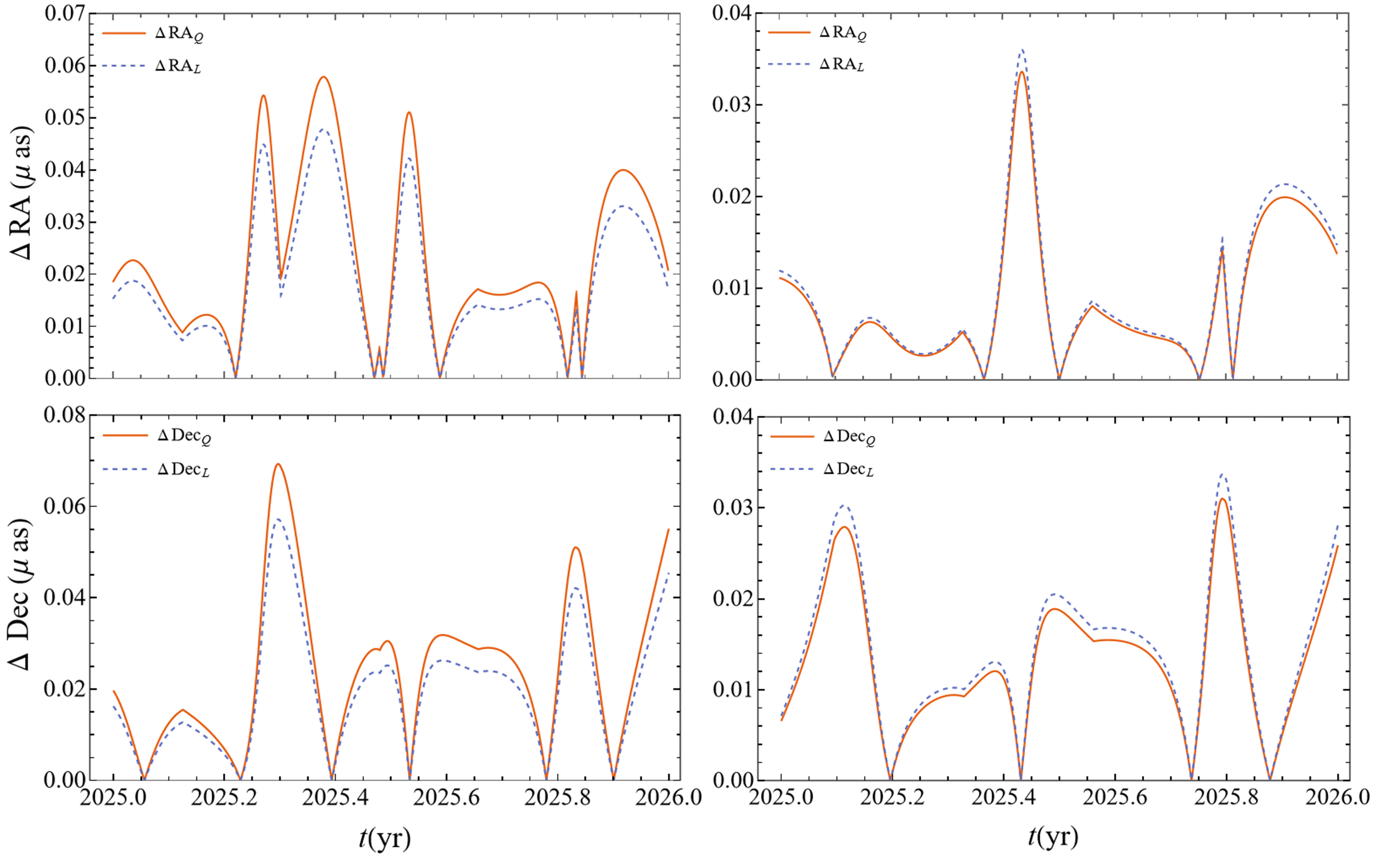}
    \caption{Difference in terms of right ascension (\textit{upper panels}) and declination (\textit{lower panels}) as a function of time between a Keplerian orbit and an orbit where contribution to the precession rate due to the quadupolar correction $\dot{\varpi}_Q$ (\textit{orange lines}) and $\dot{\varpi}_L$ (\textit{blue dashed lines}) are taken into account. The third body is an asteroidlike object orbiting near Mercury, with a semimajor axis of $a=0.5$ AU (\textit{left panels}) and $a=0.6$ AU (\textit{right panels}).}
    \label{fig:asteroid_time}
\end{figure*}

\begin{figure*}
        \includegraphics[width=1.7\columnwidth]{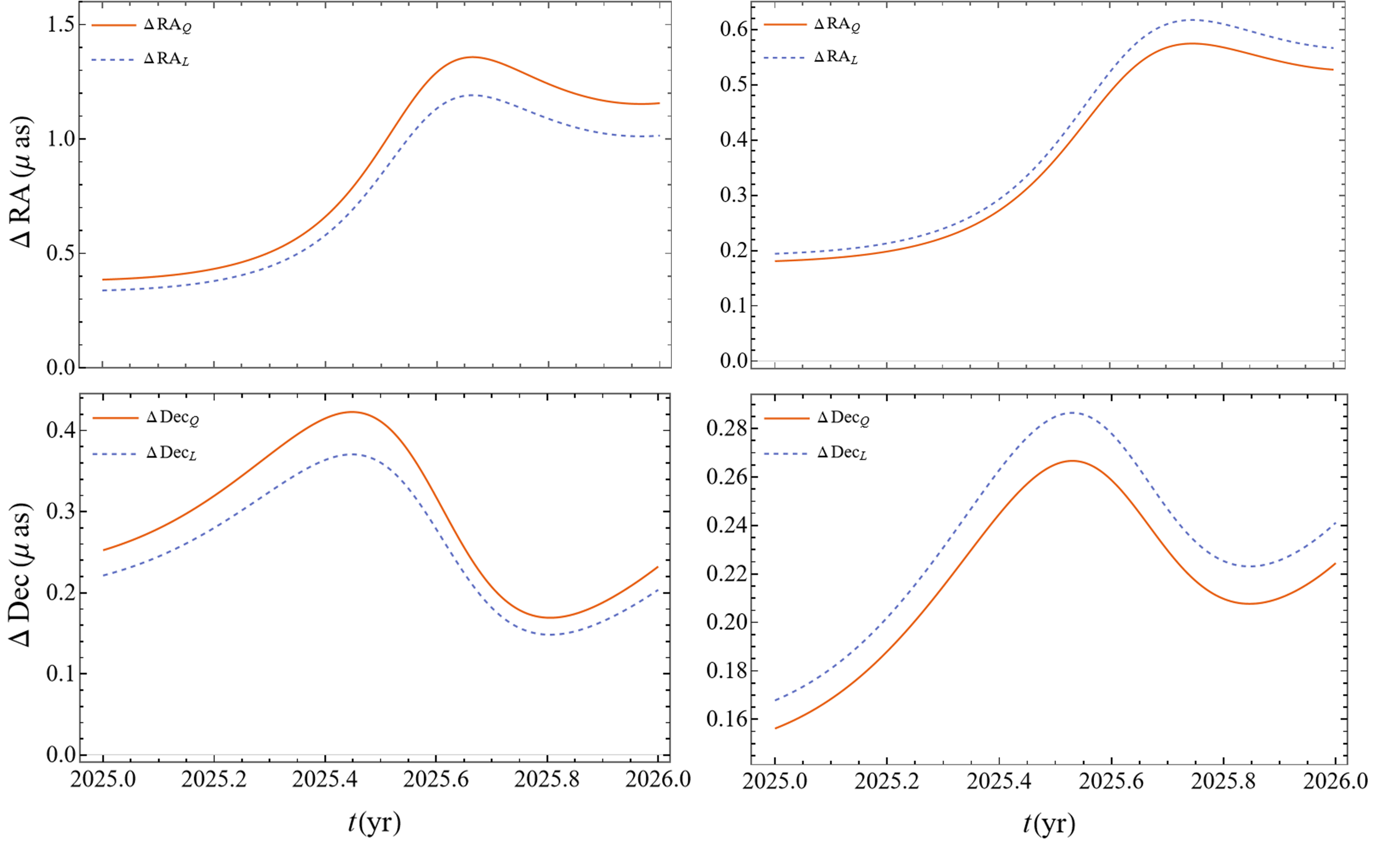}
        \label{subfig:ra jupiter 8}
    \caption{Same as in Fig. \ref{fig:asteroid_time}, but with the third object orbiting around Jupiter with a semimajor axis of $a=7$ AU (\textit{left panels}) and $a=8$ AU (\textit{right panels}).}
    \label{fig:asteroid_time2}
\end{figure*}

Finally, we study the differences in celestial coordinates as functions of time proposed in the previous subsection for the case of a putative light test particle orbiting around the Sun, such as an asteroid, whose orbit is modified considering just PN perturbative effects caused by another planet of the Solar System. We consider, separately, the cases in which such a planet is Mercury or Jupiter. The choice of these planets represents two limit cases, since Mercury has a large orbital eccentricity ($e_B=0.21$) with respect to other planets in the Solar System, and Jupiter is the most massive planet ($m_2\sim 0.001\,M_{\odot}$). In Figs. \ref{fig:asteroid_time} and \ref{fig:asteroid_time2}, we depict the contributions of the quadrupolar ($\Delta$Dec${}_Q$) and angular momentum ($\Delta$Dec${}_L$) terms to the the differences in the right ascension and the declination of the putative asteroid as orange solid and blue dashed lines, respectively. In these figures, the celestial coordinates have been calculated in the geocentric equatorial coordinate system, so we denote them by RA and Dec to distinguish them from the celestial coordinates used in the previous section. The $\Delta$ operator denotes the absolute value of the difference between the celestial coordinate in the Keplerian approximation and the orbit, taking into account the specific PN contribution. Figure \ref{fig:asteroid_time} is particularized for planet Mercury, where the left and right panels show results obtained by setting the semimajor axis of the asteroid's orbit to $a=0.5$ AU and $a=0.6$ AU, respectively. Figure \ref{fig:asteroid_time2} is particularized for Jupiter where left and right panels show results obtained by setting the semimajor axis of the asteroid's orbit to $a=7$ AU and $a=8$ AU, respectively. For all cases, the eccentricity of the asteroid's orbit is set to $e=0.2$. In general, for small $a$, we observe that the largest contribution comes from the quadrupolar term $\dot{\varpi}_Q$.  Increasing the semimajor axis of the asteroid's orbit, this contribution decreases and becomes comparable to or smaller than the one ascribable to the angular momentum. Looking at the modifications of the orbit induced by Jupiter, the contributions are one order of magnitude larger than the previous case. Also in the right panels of Fig. \ref{fig:asteroid_time2}, the contribution due to the angular momentum clearly exceeds the PN quadrupolar term due to the greater value of the semimajor axis, $a=8$ AU.  These differences could be detected by instruments in the near future, such as Theia-like satellites, which promise to reach an accuracy of tenth of $\mu$as \cite{2021ExA....51..845M}.

\section{Discussion and Conclusions}
\label{sec:discussion}

We investigated the impact of having eccentric orbits in the hierarchical and restricted three-body problem within post-Newtonian approximation. We estimated the contributions to the periastron shift of a test particle when there is a second body in an eccentric orbit around the central mass in a coplanar configuration. Differently from previous studies like \cite{yamada}, having an eccentric binary leads to some modifications in the correction terms of the averaged metric of Eq. \eqref{eqn:av_metric}, containing the eccentricity and the argument of periastron of the binary system. These additional terms modify the Newtonian and the quadrupolar corrections. In the former, the contribution of Eq. \eqref{eqn:contr2} aligns with the Newtonian quadrupolar result in hierarchiacal triplets \cite{willhex} if the limits we impose on the masses are taken into account. Regarding the relativistic corrections, the same terms found in \cite{yamada}, results to be multiplied for a function of $e_B$, $\omega_B$ and $\omega$, whose presence is ascribable to the averaging process on elliptical orbits. Comparing the different contributions, we showed that the quadrupolar term, labeled as $\dot{\varpi}_Q$, is smaller than $\dot{\varpi}_L$ when we do not consider extreme eccentricities for the binary and under the assumption $a \gg a_B$ stands. In general, the perturbative 1PN terms represent a small percentage of $\dot{\varpi}_r$ and $\dot{\varpi}_N$ for every $M$ and $q$ we have considered.

We designed several different astrophysical systems to estimate these contributions, such as planetary or SMBH systems. We noted that the Newtonian contribution tends to decrease when the central mass grows, depending on the mass ratio $m_2/M$, while others are affected by $m_2$ only. Moreover, when considering compact objects, the parameter space must be explored carefully, since only orbital elements that give rise to orbital periods on timescales below the gravitational wave damping timescale can be considered. We applied our model to the Galactic Center star S87, in order to study whether the presence of an IMBH in the Galactic Center would give detectable effects on its periastron shift. We showed that the differences in the right ascension and the declination for this object may be observable with GRAVITY in the near future, depending on the combination of the orbital parameters of the central binary. In the case of an asteroid in the Solar System, we considered modifications induced by Mercury and Jupiter at the 1PN level, and showed that the differences in the right ascension and the declination could be detected by a Theia-like satellite.

Summarizing, we have computed the 1PN contributions to the periastron advance of a test particle orbiting around an eccentric binary system.  Although our calculations are based on several simplifying assumptions, we were able to show the potential of our method in different astrophysical scenarios. More accurate modeling is needed to allow us to apply these results to any hierarchical three-body system, such as one with nonzero inclination between the orbits of the binary system and the third body, and including secular variations of the other orbital elements, which will be the subject of future applications.

\section*{Acknowledgments}
I.D.M. and R.D.M. acknowledge financial support from the Grant No. PID2021-122938NB-I00 funded by MCIN/AEI/10.13039/501100011033. R.D.M. also acknowledges support from Consejeria de Educaci\'on de la Junta de Castilla y Le\'on and the European Social Fund.  I.D.M. also acknowledges support from the Grant No. SA097P24 funded by Junta de Castilla y Le\'on and by "ERDF A way of making Europe". M.D.L. and P.F. acknowledge the support of Istituto Nazionale di Fisica Nucleare (INFN) {\it iniziativa specifica} TEONGRAV.

\bibliographystyle{apsrev4-1}
\bibliography{biblio}
\end{document}